\documentclass{PoS}

\usepackage[T1]{fontenc}
\usepackage[utf8]{inputenc}
\usepackage{amsmath}
\usepackage{slashed}
\usepackage{ulem}

\newcommand{\MS}{{\overline{{\rm MS}}}}
\newcommand{\MEV}{{\rm MeV}}
\newcommand{\GEV}{{\rm GeV}}

\title{Quark Masses from Lattice QCD}

\ShortTitle{Quark Masses}

\author{\speaker{Francesco Sanfilippo}\\
  School of Physics and Astronomy, University of Southampton,\\
  SO17 1BJ Southampton, United Kindgdom\\
  E-mail: \email{fr.sanfilippo@gmail.com}}

\abstract{In this talk I review several topics concerning the
  determination of quark masses by means of lattice QCD simulations,
  with particular focus on recently introduced techniques of
  non-perturbative renormalisation, the determination of heavy quark
  masses, and the proper quantification of the source of isospin
  breaking effects in the light quark sector.}

\FullConference{The 32nd International Symposium on Lattice Field Theory,\\
		23-28 June, 2014\\
		Columbia University New York, NY}

\begin{document}

\section{Introduction}

Quark masses are fundamental parameters appearing in the QCD
lagrangian, and as such they cannot be predicted by the
theory. Instead, they are extracted from a comparison of theoretical
expressions with experimentally measured values of various physical
observables. Theories beyond the Standard Model, aiming at the grand
unification of fundamental interactions and based on a specific gauge
group, also predict the symmetry pattern amongst various Yukawa
couplings which then translates into equalities amongst ratios of
various fermion masses. The fundamental gauge group is then assumed to
be acceptable as long as the theoretical ratios among quark/lepton
masses agree with actual values. It is therefore of fundamental
importance to reliably define and accurately determine the quark
masses. On the practical side, quark masses decisively enter the
theoretical expressions for a vast number of physical processes so
that their more accurate values significantly improve the theoretical
precision.

In that respect lattice QCD has played the essential role over the
past two decades, and nowadays the values of the quark mass and
$\alpha_s(\mu)$ are more and more dominated by the results from
numerical simulations on the lattice.~\footnote{This does not apply to
  top quark, which will not be reviewed here}

It is of crucial importance to properly define the quark mass. In
pertutbation field theory the mass can be defined as a pole of the
propagator. Such a definition, known as the pole quark mass, is a
purely perturbative concept and suffers from infrared ambiguities
known as renormalons. Non-perturbatively, however, the quark pole mass
cannot be defined since the renormalised quark propagator has no
distinct pole due to confinement in QCD.

The purpose of this review is to present an updated comparison of
different determinations of the quark masses computed by means of
numerical simulations of QCD on the lattice. As far as the light quark
masses are concerned, most of the material has already been discussed
in great detail in the FLAG review~\cite{Aoki:2013ldr}. Here we only
update those results and focus on a discussion of the mass difference
$m_d-m_u$, for which results have been recently published. Most of
this review is devoted to the lattice QCD determination of the heavy
quark masses, a subject that has not been discussed in
Ref.~\cite{Aoki:2013ldr}.

The values of bare parameters in the Lagrangian depend on the adopted
regularisation. A meaningful comparison of various determinations of
the quark mass values can be made only in a specific renormalisation
scheme and at a specific renormalisation scale. It is customary to
quote quark masses in the $\MS$ scheme. The referential scale is
$\mu=2$~GeV, which is expected to be large enough to make a
parturbative matching between the $\MS$ and a non-perturbative
renormalization scheme (suitable for lattice studies) reliable, and to
be small enough with respect to the hard lattice cut-off scale. With
more and more simulations performed at very fine lattice spacings the
referential scale is often pushed to $\mu=3$~GeV and will be pushed
even higher. One should also stress that the choice of $\MS$ as a
reference renormalisation scheme is also made when comparing lattice
results obtained by using different regularization schemes. That step
is still widely adopted but is in principle unnecessary since the
computation of the mass renormalization constant is made in
renormalization schemes suitable for lattice QCD computations, such as
RI-MOM~\cite{Martinelli:1994ty}, SF~\cite{Luscher:1992an},
RI-SMOM~\cite{Sturm:2009kb}, but not $\MS$, which is inherently
perturbative.

In addition to the FLAG review released in 2013~\cite{Aoki:2013ldr},
the Particle Data Group recently updated a section dedicated to the
quark masses~\cite{Agashe:2014kda}. In that latter review the results
obtained by using different approaches (sum rules, effective theories,
lattices, etc.) have been combined and the quoted averages are shown
to be heavily dominated by the results obtained from the studies of
QCD on the lattice.

The remainder of the present review is organised as follows: In
Sec.~\ref{renormalisation} we contextualise the determination of quark
mass in the more general framework of renormalisation of QCD, describe
a commonly used approach to determine quark masses and stress the
importance of quark mass ratios; we will devote special attention to
the additional problems arising when defining a mass indepentent
renormalisation scheme in $n_f=2+1+1$ simulations, that are becoming
more an more available. In Sec.~\ref{other_inputs} we discuss results
obtained by using less common inputs to compute the quark masses on
the lattice. In Sec.~\ref{moments} we discuss an alternative method of
computing the quark masses through the moments of correlation
functions. Sec.~\ref{b_quark} is devoted to various determinations of
the $b$ quark mass, while in Sec.~\ref{ud_diff} we discuss the recent
works in which the $d$-$u$ quark mass difference has been computed. We
conclude in Sec.~\ref{conclusions}.

\section{Renormalisation}\label{renormalisation}

The computation of quark masses starts by defining a procedure to tune
bare parameters in such a way as to keep the physics unchanged while
removing the cut-off. The lines of constant physics, described by such
a set of parameters, are constrained to describe the real world in the
continuum limit. A common procedure consists in reproducing the mass
of the lightest pseudo-Goldstone bosons, and to use an additional
quantity (such as the mass of $\Omega$, or a pion/kaon decay constant)
to determine the lattice spacing. Light pseudoscalar meson masses are
highly sensitive to the light quark masses, they are easy to compute
on the lattice even with a limited numerical effort, and their
dependence on quark masses is well described by the Chiral
Perturbation Theory which can also be used to efficiently correct for
the (exponentially small) finite volume effects of the lattice.

After tuning the quark masses to reproduce the physical pseudoscalar
meson masses, the lattice regularised theory is renormalised, but to
make sense of the bare parameters in the Lagrangian, further steps are
needed. A renormalised quark mass can be introduced by means of the
chiral vector Ward identity, which ensures that the product of the
quark mass and the corresponding scalar density ($S=\bar q q$) is
Renormalisation Group Invariant (RGI), and therefore $Z_m=1/Z_S$. In
other words, in any renormalisation scheme that does not break the
chiral symmetry,
\begin{equation}
  m_q^{ren}(\mu) = Z_m(\mu) m_q^{bare} = m_q^{bare}/Z_S(\mu)\,. \label{mqms}
\end{equation}

Such relations\footnote{If, instead, the adopted lattice
  regularisation breaks the chiral symmetry, in Eq.~(\ref{mqms}) an
  additive renormalisation on the right hand side is needed as well,
  namely, $m_q^{ren}(\mu) = Z_m(\mu) [m_q^{bare}-m_{cr}]$, with
  $m_{cr}$ is a critical mass, i.e. the amount of explicit chiral
  symmetry breaking by the lattice regulator.} allow for a meaningful
notion of the renormalised quark mass at a fully non-perturbative
level, in schemes such as RI-MOM, RI-SMOM or SF.  Relations to the
reference $\MS$ renormalisation scheme can be established in
perturbation theory. It is then a question which scale can be probed
by the non-perturbative method, and then to assess whether or not that
scale is high enough to keep the uncertainties due to truncation in
perturbative expansion small.

For example, the Rome-Southampton Method (RI-(S)MOM) to compute the
renormalisation constants non-perturbatively requires the existence of
a window $1/L\ll\Lambda_{QCD}\ll\mu\ll \pi/a$, with $L$ being the
length of a side of the lattice box, in which the uncertainties
related to the finite cut-off and to the perturbative matching to
other renormalisation schemes are controllable and small.  In practice
this requirement translates to the condition that the lattice spacings
should be smaller than about 0.1\,fm ($a^{-1}\simeq2$\,GeV). In the
case of staggered quark action, however, additional complications
prevented so far a simple implementation of the procedure.

The scaling window needed in the RI-(S)MOM is not needed in the method
based on the use of SF where a series of additional simulations need
to be made with special boundary conditions on a box of finite
extension $T$, which ensure the renormalisation scale to be $\mu=1/T$.

When non-perturbative renormalisation methods cannot be easily
applied, one can rely on the QCD perturbation theory to compute
$Z_m(\mu)$. Unfortunately, however, perturbative calculations on the
lattice become rapidly too complex with an increase in the order of
the perturbative expansion, which in practice means that going beyond
two loops becomes extremely difficult to do (cf. the three loop
stochastic computation presented by M.Brambilla at this
conference~\cite{Brambilla:2014bja}).

In Sec.~\ref{moments} we will discuss a sophisticated approach that
allows to make use of high-order continuum perturbation theory.
Before discussing in detail the renormalisation of $n_f=2+1+1$
simulations, we comment extensively on the physical content of ratios
of bare quark masses.

\subsection{RGI ratios of quark masses}\label{sec_ratio}

\begin{figure}
  \includegraphics[viewport=30bp 125bp 592bp 650bp,clip,width=0.47\textwidth]{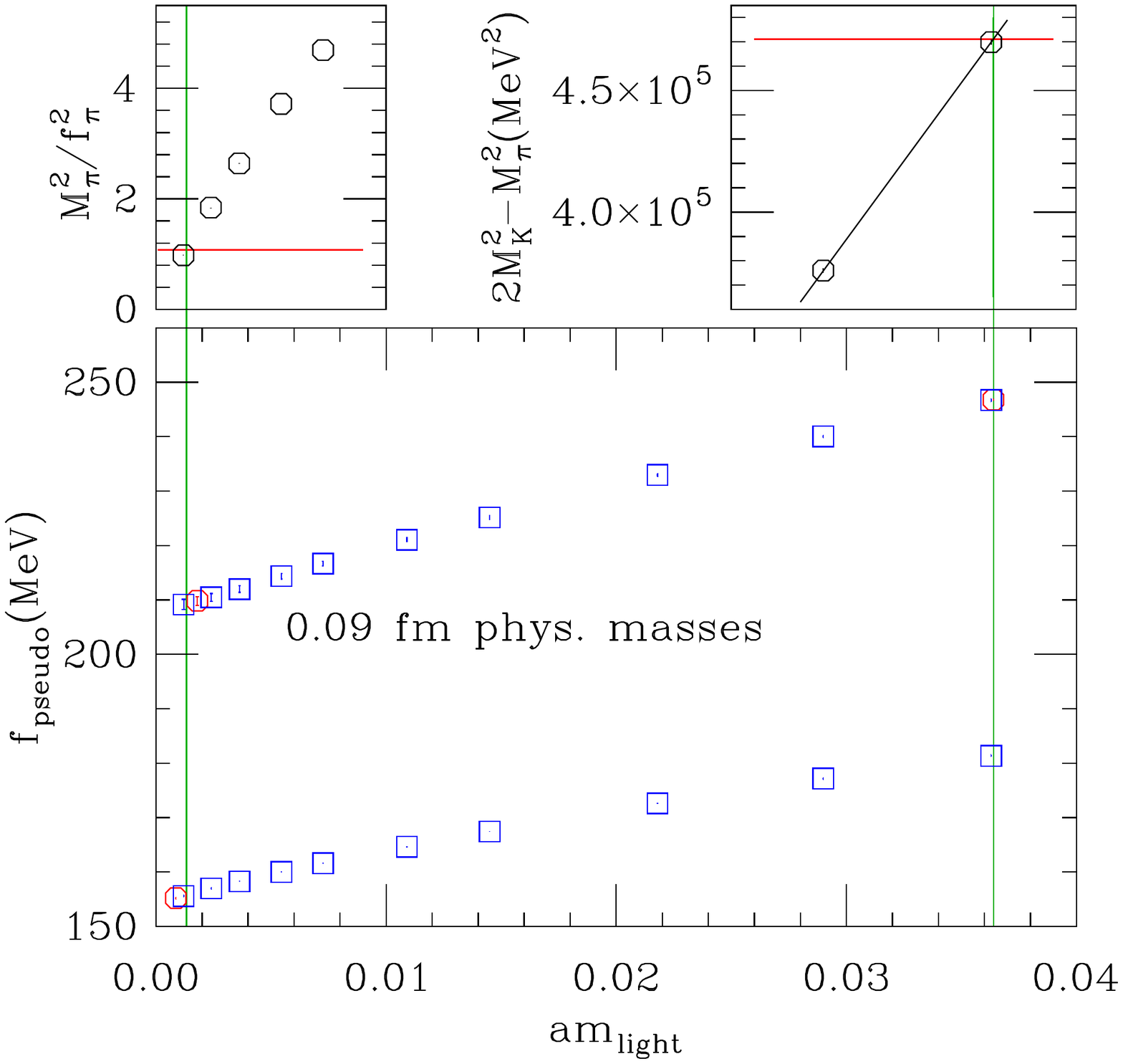}
  \includegraphics[viewport=30bp 120bp 612bp 692bp,clip,width=0.48\textwidth]{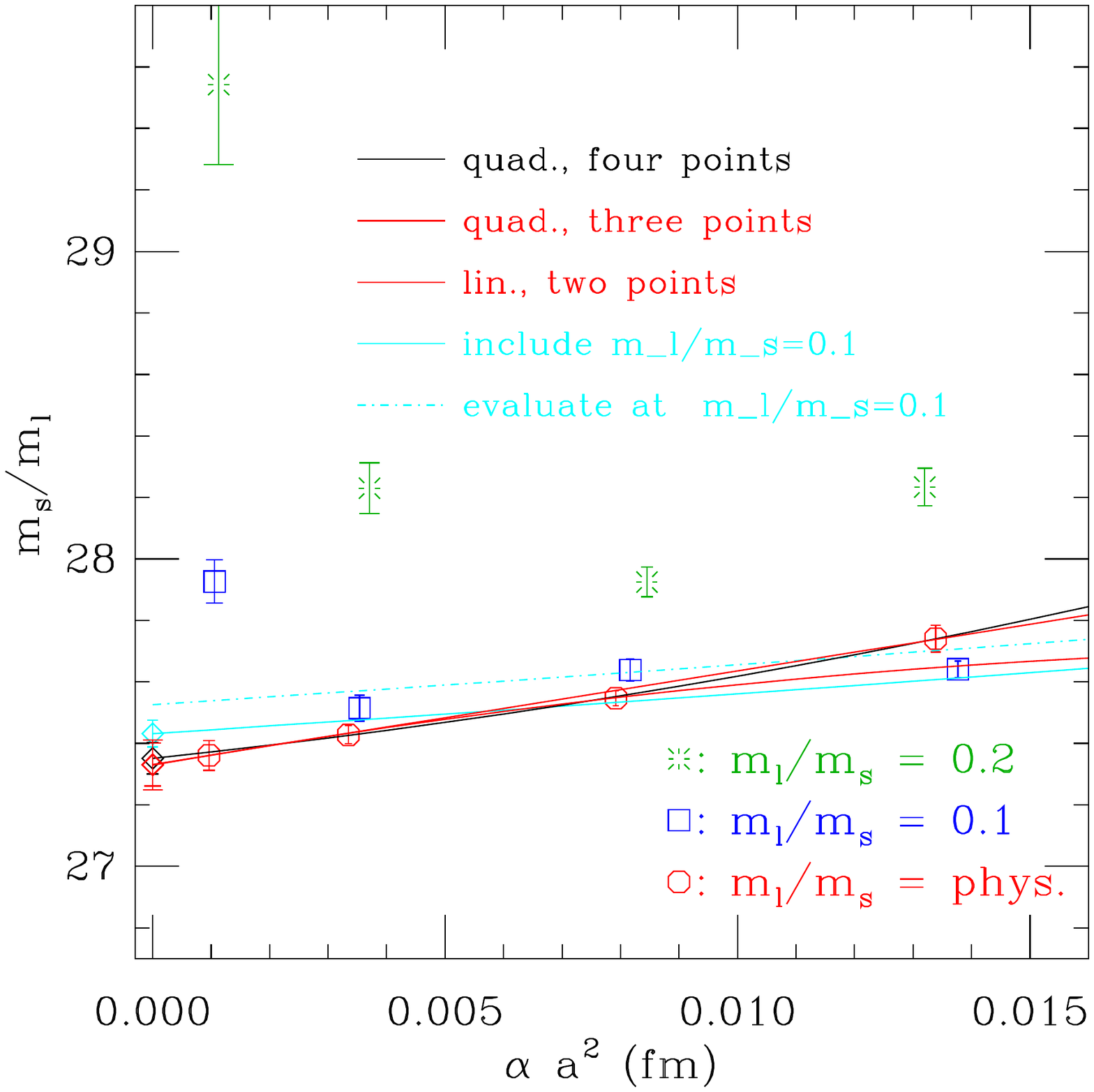}
  \caption{\uline{Left panel}: tuning of the light and strange quark
    masses at a fixed lattice spacing (0.09 fm); different points
    corresponds to mesons simulated numerically, with quarks of common
    mass $m_{l}$ (x-axis), where vertical green lines show the value of
    $m_{l}$ at which the interpolation of $M^2_\pi/f^2_\pi$ (top-left
    inset), $2M^2_K-M^2_\pi$ (top-right) and the decay constant $f$
    (bottom) reproduce their physical value (red horizontal
    lines). \uline{Right panel}: continuum extrapolation of the ratio
    computed at various lattice spacings. Plots
    taken from Ref.~\cite{Bazavov:2014wgs}.}\label{milc}
\end{figure}

The knowledge of $am^{bare}(a)$ is already of invaluable
importance. If the regularisation preserves the chiral symmetry (or
its $U(1)_A$ remnant), the quark masses renormalises multiplicatively,
so that in the mass independent renormalisation schemes ($\partial
Z_{m}/\partial m_q =0$) the ratio of renormalised quark masses is
equal to the ratio of bare quark masses:
\begin{equation}
  \frac{m_{q_{1}}^{ren}}{m_{q_{2}}^{ren}}=\frac{am_{q_{1}}^{bare}}{am_{q_{2}}^{bare}}\,.
\end{equation}
This relation holds as long as QED is not included in the
simulation. The above conditions are respected (for example) in the
computations made by the Fermibalb/MILC
collaboration~\cite{Bazavov:2014wgs} (cf. also contribution by
J.~Komijani at this conference) based on their recent $n_f=2+1+1$ HISQ
regularised quark simulations. To determine the ratio between the
strange quark mass $m_s$ and the average of up and down quark masses,
$m_l=(m_u+m_d)/2$, they proceeded in two steps. First, at each value
of the lattice spacing they tuned the bare light quark masses in such
a way that $\frac{\left(a^{2}\right)M_{\pi}^{2}\left(am\right)}
{\left(a^{2}\right)f_{\pi}^{2}\left(am\right)}$ computed on the
lattice reproduces the experimentally measured
$\left(M_{\pi}/f_{\pi}\right)_{exp}^{2}$. The resulting
$am_{l}^{bare}$ is then converted to physical units after using the
lattice spacing $a$ extracted from
$af_{\pi}\left(am_{light}^{bare}\right) = f_{\pi}^{exp}$. The bare
strange quark mass, instead, is fixed by tuning
$M_{ss}^2=2M_{K}^{2}-M_{\pi}^{2}$ to its physical value (see left
panel of Fig.~\ref{milc}). To LO in ChPT $M_{ss}^2$, is independent on
$m_l$, so the tuning of $m_s$ almost decouples from that of
$m_l$. With such a determined value of $m_s$, the charm quark mass
$m_{c}$ is obtained from the requirement that the mass of the
heavy-strange pseudoscalar meson coincides with $M_{D_{s}}^{exp}$.

The ratios $m_{s}/m_{l}$ and $m_c/m_s$ are computed at each lattice
spacing, and are then extrapolated to the continuum limit as shown in
the right panel of Fig.~\ref{milc} resulting in
\begin{eqnarray}
  m_s/m_l & = & 27.35 (5) (^{+10}_{-7})\,, \\
  m_c/m_s & = & 11.747 (19) (^{+59}_{-43})\,.
\end{eqnarray}
These are the most accurate results for the quark mass ratios to date.

\subsection{Renormalisation constants in $\boldsymbol{n_{f}=2+1+1}$ simulations}\label{reno_massive_qua}

The ETM collaboration was the first to produce results for quark
masses from simulations including $N_{f}=2+1+1$ dynamical quark
flavours. Results presented in Ref.~\cite{Carrasco:2014cwa} were
obtained from the analysis along the lines of their previous work with
$N_f=2$~\cite{Blossier:2010cr}. They computed non-perturbatively the
mass renormalisation constants in the RI-MOM renormalisation scheme
and performed matching to the $\MS$ renormalisation scheme by using
the expressions derived to four loops in perturbation
theory~\cite{chetyrkin}. In order to define a mass-independent
renormalisation scheme, the renormalisation scale $\mu$ must be much
larger than all the other scales of the theory, and, in particular, we
must have $\mu\gg m$, with $m$ the mass of the heaviest quark. In
RI-MOM scheme this is achieved by defining renormalisation constants
at the chiral point.

In the case of their previous $n_f=2$ set of simulations, ETM
collaboration computed renormalisation constants on the same set of
gauge configurations used to measure hadronic
observables. Extrapolating to the chiral limit in that case comes as a
virtue out of the necessity to simulate a wide range of non-physical
sea pion masses (280-500 MeV) needed for controlled extrapolation of
observables to the physical point at which the sea pion mass is
$m_\pi^{phys}$.

Conversely, results obtained from simulations carried out around the
physical charm quark mass in the new $n_f=2+1+1$ simulations cannot be
safely extrapolated to the chiral limit of the four quark
masses. Therefore the renormalisation constants cannot be computed
directly on $n_f=2+1+1$ configurations on which bare quark masses are
computed.~\footnote{Renormalisation constants computed at the physical
  charm quark mass would define a mass-dependent renormalisation
  scheme. Though perfectly legitimate in principle, this additional
  mass dependency introduces additional complications when matching to
  $\MS$ scheme, and therefore is far from be welcome.} To overcome
this problem, the ETM collaboration specifically generated a set of
gauge configurations with $n_f=4$ mass degenerate light quarks. The
mass of the dynamical quarks is varied as to allow a continuum
extrapolation~\cite{Dimopoulos:2011wz} of the renormalisation
constants. With these ingredients at hand the average up-down, strange
and the charm quark masses were found to be:
\begin{eqnarray}
 m_l^\MS(2\,\GEV) & = & 3.70(17)\,\MEV\,, \\
 m_s^\MS(2\,\GEV) & = & 99.6(4.3)\,\MEV\,,\label{msetm}\\
 m_c^\MS(m_c) & = & 1.348(46)\,\GEV\,.\label{mcetm}
\end{eqnarray}
The authors also noted the advantage of working with ratios of similar
quantities (eg. $M_{\pi}/M_{\bar{s}s}$, $M_{D_{s}}/M_{\bar{c}^\prime s}$),
because they are less sensitive to cut-off effects and therefore
their continuum extrapolation is much smoother.

Another possibility to define a mass independent scheme is to use the
Schr\"odinger Functional (SF) renormalisation procedure, where
simulations can be carried out directly in the massless theory, thus
avoiding the need to extrapolate to the chiral limit. In the theory in
which the heavy quarks $m_h$ are included, one can use the Step
Scaling approach to evolve the renormalisation constants computed with
physical (massive) heavy quarks up to a scale $\mu \gg m_h$, so that
the terms $\propto (m_h/\mu)^n$ can be safely neglected. Such an
approach is currently being investigated by the RBC/UKQCD
collaboration (cf. J.~Frison contribution,
Ref.~\cite{Frison:2014esa}).

\section{Other physical inputs}\label{other_inputs}

\subsection{Fixing the quark masses through the baryon spectrum}

\begin{figure}
  \includegraphics[width=0.47\textwidth]{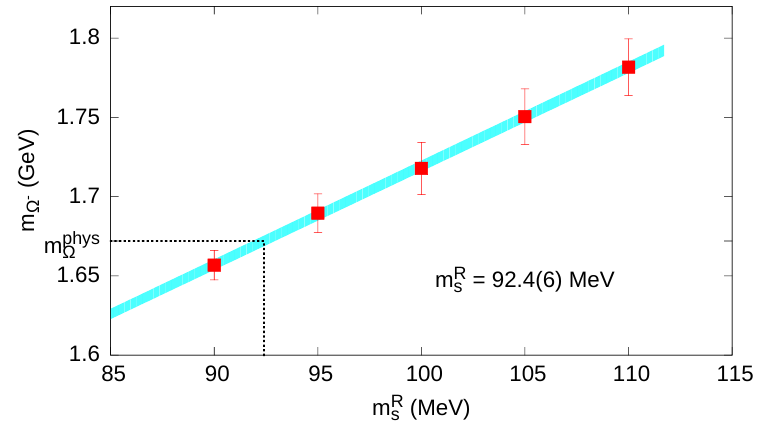}
  \includegraphics[width=0.47\textwidth]{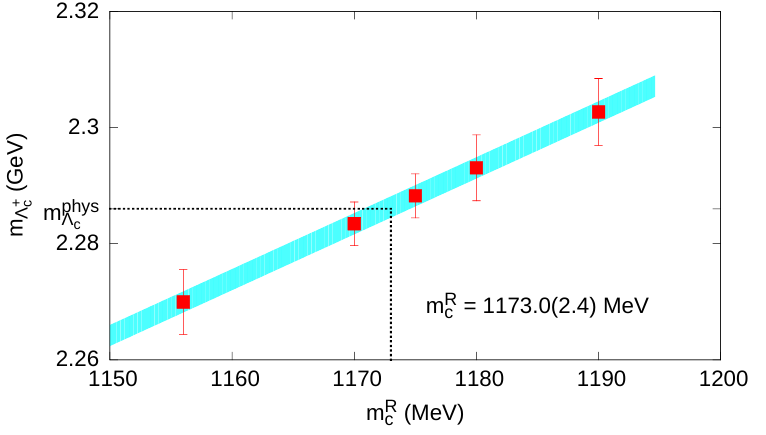}
  \caption{Interpolation of $\Omega^-$ (left panel) and $\Lambda_c^+$
    (right panel) baryons respectively to physical renormalised
    strange and charm quark mass. Plots taken
    from Ref.~\cite{Alexandrou:2014sha}.}\label{dina}
\end{figure}

Of all the possible observables used to compare theory and experiments
to extract the values of physical quark masses, the most commonly used
are the pseudoscalar meson masses, mainly because they are the
simplest physical quantities to compute on the lattice and the
corresponding results are very accurate. Nonetheless it is possible to
use other quantities. For example, the ETM collaboration has recently
reported on their results for the strange and charm quark masses by
computing the baryon masses on the same set of $n_f=2+1+1$ gauge
configurations already discussed above, in Sec.~\ref{reno_massive_qua}
(see also the contribution by Ch.Kallidonis at this
conference~\cite{Alexandrou:2014sha}).

More specifically, they fixed $m_{s}$ from $\Omega$, the valence
configuration of which is $sss$, and $m_{c}$ from the singly charmed
baryon $\Lambda_{c}$. By linearly interpolating the baryon masses
expressed in terms of $m_s^{ren}$ and $m_c^{ren}$ to their physical
values, as illustrated in Fig.~\ref{dina}, they were able to extract
the quark masses, while the lattice spacing has been determined by
using the pion and/or the proton mass. The continuum and chiral
extrapolations are made by relying on an empirical ansatz,
\begin{eqnarray}
  M_{\Omega} & = & M_{\Omega}^{chir}+c_{\Omega}M_{\pi}^{2}+d_{\Omega}a^{2}\,,\\
  M_{\Lambda_{c}} & = &
  M_{\Lambda_{c}}^{chir}+c_{\Omega}^{\left(2\right)}
  M_{\pi}^{2}+c_{\Omega}^{\left(3\right)}M_{\pi}^{3}+d_{\Omega}a^{2}\,.
\end{eqnarray} 
Lacking a solid effective theory framework to perform chiral
extrapolation and to reliably estimate the finite volume effects, this
analysis is more challenging than those based on using the
pseudoscalar mesons, mainly because of the difficulties in assessing
the size of systematic errors. The resulting strange and charm quark
masses,
\begin{eqnarray}
  m_s^\MS(2\,{\rm GeV}) & = & 92.4(0.6)(2.0)\, {\rm MeV}\label{dina1} \\
  m_c^\MS(2\,{\rm GeV}) & = & 1.173(2)(17)\, {\rm GeV}\,,\label{dina2}
\end{eqnarray}
appear to be in reasonable agreement with values obtained when using
mesons as inputs, Eqs.~(\ref{msetm},\,\ref{mcetm}), which is an
encouraging consistency check. We reiterate that the systematic error
estimates in Eqs.~(\ref{dina1},\,\ref{dina2}) are presumably less robust
than those presented in Eqs.~(\ref{msetm},\,\ref{mcetm}).

\subsection{Global fit approach}

It is a common practice to determine the quark masses in dedicated
lattice QCD analysis, by choosing a minimal number of inputs to
renormalise the theory, and then to use the obtained quark mass values
in calculations of other physical quantities.

Recently the RBC/UKQCD collaboration has presented a large set of
results obtained from simulations performed around the physical pion
mass~\cite{Blum:2014tka}. In order to correct for the (small)
difference between the simulated quark mass and the physical one they
made a very short $\mathcal{O}(3\%)$ chiral extrapolation in which
they combined the results of simulations made near the physical pion
mass with those corresponding to heavier pions. In addition to the
three experimental inputs used to tune the physical light quark masses
($m_\pi$, $m_K$ and $M_\Omega$), they computed several other
quantities ($f_\pi$, $f_K$, $B_K$, \dots) and then performed a
simultaneous fit of all the data by using the expressions derived in
chiral perturbation theory in which they were able to further monitor
the dependence of each physical quantity on the sea quark mass. Such a
dependence further constrains the quark masses, the values of which
turn out to be extremely accurate, namely,
\begin{eqnarray}
  m_l^\MS(3\,{\rm GeV}) & = & 2.997(49)\, {\rm MeV}\\
  m_s^\MS(3\,{\rm GeV}) & = & 81.64(1.17)\, {\rm MeV}\,.
\end{eqnarray}

We notice that a global fit approach allows to determine all the
observables while keeping all correlations under control. The approach
could be in principle extended to include even more observables, but
it becomes more complicated to verify consistency of different fit
ansätze.

\section{Renormalised quark mass from the moments of correlation functions \label{moments}}

A few years ago the HPQCD group has proposed to bypass the
intermediate step non-perturbative renormalisation and obtain directly
the renormalised quark mass in the $\MS$ scheme, by computing an RGI
quantity on the lattice and matching it to its counterpart computed in
the continuum perturbation theory, expressed in terms of
the $\MS$ quark masses and the running coupling~\cite{Allison:2008xk}.

Moments of the correlation function are nowadays used in lattice QCD
for a range of different quantities. In Ref.~\cite{Allison:2008xk}
they have been used for the first time to compute the charm quark mass.

After summing up the correlation function over the space, one first
defines the moments with respect to time,
\begin{equation}
  G_{n}^{\left(j\right)}=\sum_{t}{\left(t/a\right)^{n}}G^{\left(j\right)}\left(t\right),\qquad
  G^{\left(j\right)}\left(t\right)=\left(am_{c}^{bare}\right)^{2}\sum_{\vec{x}}\left\langle
  j\left(\vec{x},t\right)j(\vec{0},0)\right\rangle\,,
\end{equation}
where $G_{n}^{\left(j\right)}$ is the $n$-th moment of the
non-perturbatively computed two-point correlation function, $j = \bar
c \Gamma c$ ($\Gamma=\gamma_5,\gamma_\mu$), with all quantities being
unrenormalised.  One then forms the convenient double ratios of
moments $G_{n}^{(j)}$ and their values $G_{n}^{(j0)}$ computed to LO
in lattice perturbation theory,
\begin{equation}
  R_{n}^{\left(j\right)}=\frac{M_{mes\,j}}{2m_{c}^{0}}\sqrt{\frac{G_{n}^{\left(j\right)}}
    {G_{n-2}^{\left(j\right)}}\frac{G_{n-2}^{\left(j0\right)}}{G_{n}^{\left(j0\right)}}}\,,
\end{equation} 
the quantities known as reduced moments. In the above expression
$M_{mes}$ is the meson mass, and $m_{c}^{0}$ is the bare mass of the charm
pole mass computed in lattice perturbation theory.  Such a mixture of
perturbative and non-perturbative quantities benefits from the
following advantages:
\begin{itemize}
  \setlength\itemsep{0.2em}
\item renormalisation constant of the operator $j$ cancel in the
  ratio, and $R_n$ is an RGI quantity;
\item cut-off effects, finite volume effects and the statistical noise
  largely cancel in the ratio;
\item the introduction of a leading order calculation of the moments
  additionally suppresses cut-off effects, so that the ratios $R_n$ can
  be smoothly extrapolated to the continuum limit.
\end{itemize}
As a last comment, we notice that the expressions for $R_n$ become
simpler if the currents used in the correlation function satisfy a
Ward Identity, because it is possible to form the RGI quantities more
easily.

Most importantly, as an effect of the known exponential decay in
space-time, moments of correlation function are largly dominated by
the short distance physics contribution, which means that for values
of $n$ not too large ($n\lesssim 10$), $R_n$ is expected to be
reliably reproduced by perturbation theory. One can deduce the values
of the quark mass $m_c^\MS (\mu)$ and of the running coupling
$\alpha_s^\MS (\mu)$ by comparing this result with the continuum
perturbation theory prediction,
\begin{equation}
  R_{n}^{PQCD}=\frac{r_{n}^{PQCD}\left(\alpha_\MS,\mu/m_{c}\right)}{2m_{c}^\MS
    \left(\mu\right)/M_{mes\,j}}\,,\label{RPQCD}
\end{equation}
where $r_{n}^{PQCD}$ has been derived at three
loop~\cite{Chetyrkin:2006xg}.

\begin{figure}
  \includegraphics[viewport=0bp 10bp 253bp 188bp,width=0.45\textwidth]{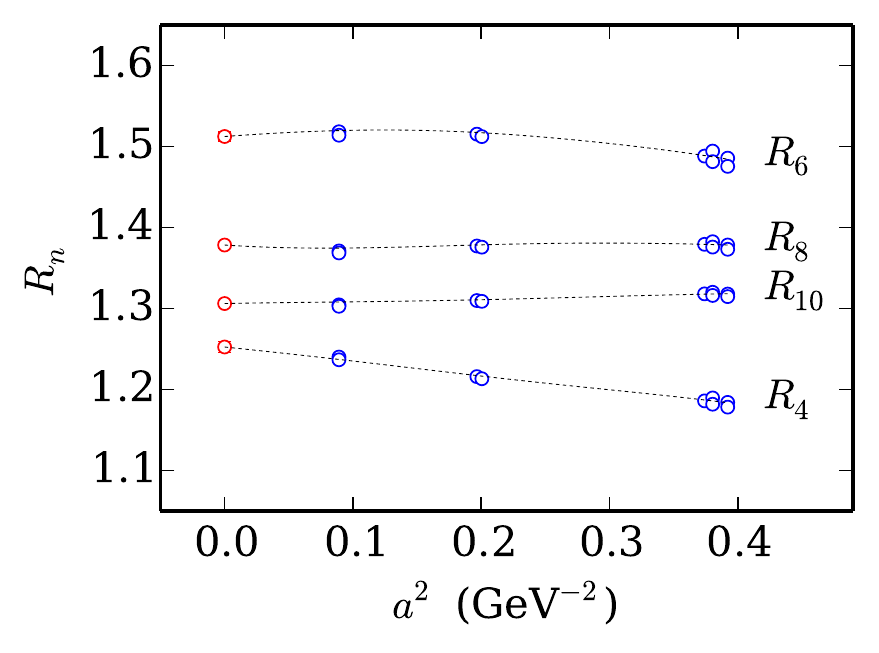}
  \includegraphics[width=0.49\textwidth]{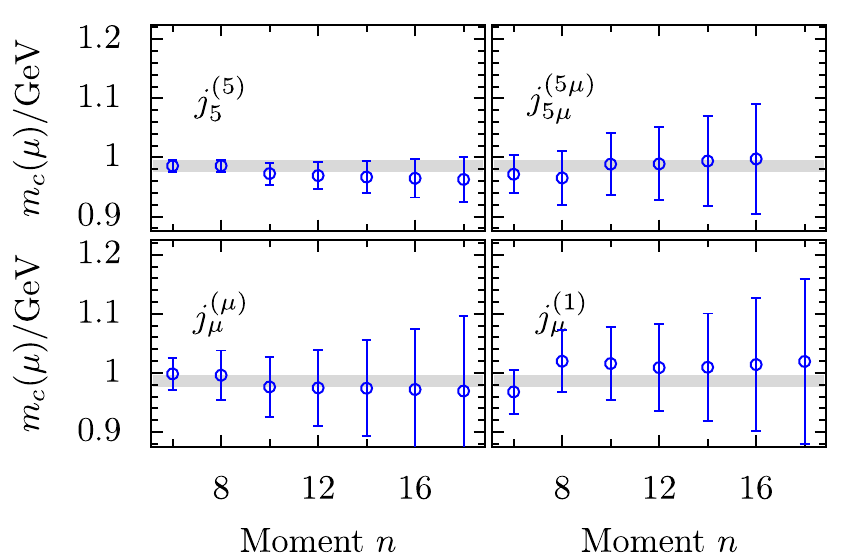}
  \caption{Left panel: continuum extrapolation of ratios of different
    order (plot taken from Ref.~\cite{Chakraborty:2014aca}).
    \uline{Right panel:} $mc(\mu)$, for $\mu$ = 3 GeV and $n_f$ = 4
    flavors, from moments of correlators built from four different
    lattice operators (grey band marking the final
    results).}\label{umcneille}
\end{figure}

In the left panel of Fig.~\ref{umcneille} we show the extrapolation to
the continuum limit of the reduced moments $R_{4,6,8,10}$ computed
numerically by the HPQCD collaboration, after tuning the bare quark
mass $am_{c}^{bare}$ in such a way as to reproduce $M_{\eta_{c}}$. In
the right panel is shown the result of $m_c(\mu)$ obtained from
comparison of the lattice results with the expression~(\ref{RPQCD}),
for several types of currents and for a wide range of moments.

The approach proposed by HPQCD collaboration is inspired by the
quark-hadron duality sum rules in which the moments of correlation
function are estimated non-perturbatively by connecting the
experimental data for differential electron-positron annihilation
cross section, $d\sigma(e^+e^- \to\gamma^\ast\to X)/ds$, with
perturbative QCD computations, in order to extract the heavy quark
mass and the running coupling. In the method proposed here, instead of
using the experimental data one computes the moments directly on the
lattice and by using various Dirac structures.

HPQCD collaboration performed various tests to check the stability of
$m_{c}\left(\mu\right)$ with respect to the variation of $n$, and the
compatibility of the results obtained from different correlation
functions. The analytic parametrisation of Eq.~(\ref{RPQCD}) of
$R_{n}$ has also been extended by including the leading power
correction, proportional to the gluon condensate, in order to try and
quantify the non-pertubative effects. Those effects are, however,
found to be negligible.~\footnote{Given the intrinsic ambiguity of the
  non-perturbative interpretation of the gluon condensate, this
  inclusion cannot be used to improve the reliability of the
  determination of the quark mass, but just as a mean to estimate the
  applicability of perturbation theory to such a computation.}

In their recent paper~\cite{Chakraborty:2014aca} the HPQCD
collaboration reported on their calculation of the charm quark mass
obtained on the set of $n_f=2+1+1$ gauge field configurations produced
by the MILC collaboration in which the improved (HISQ) regularisation
of the sea quarks has been implemented. In other words, in their
calculation the effects of the charm quark on the polarization of the
QCD vacuum have been taken into account, while the light quark mass is
varied down to the physical pion mass. By introducing a simplified
version of the ratios, and after setting the scale by using $w_0$
instead of $r_1$, they obtained, $m_c(3\,{\rm GeV},{
  n_f}=4)=0.9851(63)\,{\rm GeV}$, in good agreement with their
previous result (cf. Ref.~\cite{McNeile:2010ji}) $m_c(3\,{\rm GeV},{
  n_f}=3)=0.986(6)\,{\rm GeV}$\,. Combining the renormalised value of
the charm quark mass with the updated determination of the
charm/strange quark mass ratio, $m_c/m_s=11.652(65)$, they also
provided an accurate estimate of the strange quark mass, $m_s(2\,{\rm
  GeV},n_f=3)=93.6(8)\,{\rm MeV}$ and of the $b$ quark mass,
$m_b^\MS(m_b,\,n_f=5)=4.162(48)\,\GEV$. This whole approach can be
interpreted as a way to compute the quark mass renormalisation
constant, $Z_m^\MS(1/a)=m_c^\MS(1/a)/am_{c}^{bare}/a$, by using
$M_{mes\,j}$ as a physical input.

Results presented by the HPQCD collaboration are very accurate, and a
natural question arises: are the moments of correlation functions a
viable tool to increase the accuracy of lattice QCD determinations? So
far, only the ETM collaboration implemented the method of moments in
the computation by using the gauge field configurations with $n_f=2$
dynamical light quarks~\cite{Petschlies:2014ima}. Their preliminary
results, $m_c(3\,{\rm GeV}, n_f=3)=\{0.979(9),\,0.998(14)\}\,{\rm
  GeV}$, based on two different chiral extrapolation procedures, are
in good agreement with the values reported by the HPQCD collaboration,
as well as with the previous result of ETMC obtained by using the
standard method on the same set of gauge field configurations and by
implementing the non-perturbative RI-MOM renormalisation
program~\cite{Blossier:2010cr}. Given the lack of a comprehensive
study of systematic errors, it remains difficult to estimate the
improvement induced by the use of ratios of moments of correlation
functions instead of the ordinary non-perturbative renormalisation,
nor is it possible to accurately verify the compatibility of the
results and assess the ultimate question concerning the overall
control of non-perturbative effects of the method. One can only be
comforted by the circumstantial evidence of the smallness of the gluon
condensate contribution detected in~\cite{Chakraborty:2014aca},
suggesting that the ambiguity in separation of the perturbative and
non-perturbative effects affects the results well below the current
precision.

In conclusion, moments of correlation functions have not yet been
shown to be clearly superior to ordinary methods to define
renormalised quark mass, but are certainly a key factor to allow a
precise determination in those lattice frameworks where setting up a
non-perturbative renormalisation program is notoriously difficult (for
example in the staggered regularisation).

\section{$b$ quark mass}\label{b_quark}

Typical lattice spacings used in current lattice QCD simulations are
larger than (or of the same order as) the scale involved in the
physics of $b$ quark. For example $M_{B^+}=5.279\,\GEV$ while a
typical range of (inverse) lattice spacings of the lattice simulations
is $[2\div 4]$ GeV. For this reason specific methods dedicated
to treatment of the $b$-quark on the lattice have been designed. In
the following we describe modern approaches used to determine the $b$
quark mass.

\subsection{Binding energy of non-relativistic heavy meson}

Non-Relativistic QCD (NRQCD)~\cite{Bodwin:1994jh} is an effective
expansion of the QCD Lagrangian in terms of quark velocity $v$. The
framework is used to describe a series of physical quantities, ranging
from hadron spectroscopy to the hadronisation effects in decay
amplitudes. In Ref.~\cite{Gray:2005ur} it was proposed for the first
time to compute the quark mass in NRQCD through an analysis of the
binding energy in the bottomium system, namely,
\begin{equation}
M_{\Upsilon}^{exp}=2m_{b}^{pole}+\Delta M_{\Upsilon}\,,
\end{equation}
where $m_b^{pole}$ is the pole mass of $b$ quark and $\Delta
M_{\Upsilon}$ is the {\it binding energy}. Although the separation
between the two contributions is plagued by the {\it renormalon
  ambiguity}, that ambiguity cancels out when converting the pole mass
to the quark mass renormalised in the $\MS$ scheme. Therefore an
accurate computation of the binding energy could be in principle used
to obtain $m_b^\MS(\mu)$.

One first fixes the bare quark mass in the NRQCD Lagrangian, by
matching the spin averaged bottomium mass computed on the lattice with
the corresponding experimental value~\footnote{This choice is adopted
  to minimise the spin dependent systematic error induced by
  neglecting the higher order terms in the NRQCD Lagrangian}
\begin{equation}
  \overline{M_{b\bar{b}}}=a^{-1}\left(3aM_{\Upsilon}+aM_{\eta_{b}}\right)/4\,.\label{spin_ave}
\end{equation}
In NRQCD the zero of the energy is shifted so that the dispersion
relation for a meson reads:
\begin{equation}
  E(\vec{p})=E(0)+\sqrt{M+\vec{p}^2}-M\,,\label{disprel}
\end{equation}
where the rest energy $E(0)$ differs from the meson mass $M$ by an
unknown constant and thus cannot be used to directly fix the bare
quark mass. Instead, the simulated value of $M$ can be extracted by
studying the dispersion relation of mesons, as shown in
Eq.~(\ref{disprel}). The binding energy can be computed through the
relation,
\begin{equation}
  \Delta M_\Upsilon=E_\Upsilon(0)-2E_0\,,\label{bind_ene}
\end{equation}
where $E_0$ is the energy of an isolated $b$ quark, determined
perturbatively.

With these ingredients in hands, the pole quark mass can be computed
as
\begin{equation}
  m_b^{pole}=M_\Upsilon^{exp}-\left[E_\Upsilon(0)-2E_0\right]\,,\label{mb_nrqcd}
\end{equation}
which is then matched to the $\MS$ scheme by using the continuum
perturbation theory at the same order used to define the binding
energy.

NRQCD is a non-renormalisable effective theory. The subtraction
defined in Eq.~(\ref{bind_ene}) involves quantities that diverge as
$a^{-1}$ in the continuum limit. Power divergence cannot be completely
and unambiguously eliminated by means of perturbation theory, so that
even after subtracting $E_0$ the binding energy contains a not-fully
cancelled divergent term. This implies that no continuum limit of the
right-hand side of Eq.~(\ref{mb_nrqcd}) can be defined, and indeed
previous steps did not include it. In lattice NRQCD one is confined to
work in a range of lattice spacings small enough to keep the cut-off
effects are under control, but large enough to avoid the sizable
$\mathcal{O}(a^{-1})$ effects. Instead of a real continuum limit, only
a comparison of results obtained at different lattice spacings can be
made, and the difference amongst different ensembles is included in
the budget of systematic uncertainties.

In their recent work the HPQCD collaboration presented results based
on two different lattice spacings of the $n_f=2+1$ ASQtad gauge
configurations produced by the MILC collaboration and the NRQCD
action~\cite{Lee:2013mla}. The subtraction in Eq.~(\ref{bind_ene}) is
carried out to two loops using a mixture of automatised perturbation
theory (cfr. C.Monahan at lattice conference
2013~\cite{Monahan:2013dla}) and simulations performed at high
$\beta$. Their final result,
\begin{equation}
  m_b^\MS\left(m_b,\,n_f=5\right)=4.166(43) {\rm GeV}\,,
\end{equation}
is a very significant improvement of their first calculation presented
in Ref.~\cite{Gray:2005ur}, in which the subtraction of the divergence
had been carried out only to one loop.

Nonetheless it must be stressed that the ambiguity in cancellation of
the power divergence is an intrinsic feature of NRQCD that limits the
precision of the approach. When the method was proposed in
Ref.~\cite{Gray:2005ur}, it allowed to carry out the first unquenched
computation of the $b$ quark mass. Today, more reliable approaches
exist and the NRQCD results can be viewed as a consistency check of
other lattice methods.

\subsection{Matching HQET and QCD}

Heavy Quark Effective Theory (HQET) is another effective theory of QCD
based on heavy quark symmetry which provides an expansion in the
inverse heavy quark mass $1/m_h$. Contrary to NRQCD, HQET can be
matched to QCD order by order in $1/m_h$ without running into troubles
related to the subtraction of the power divergent term.
After a long effort, the Alpha collaboration
performed a fully non perturbative matching of HQET to QCD at
$\mathcal{O}\left(1/m_{h}\right)$~\cite{Blossier:2012qu},
\begin{equation}
  \mathcal{L}^{HQET}=\bar{\psi}_{h}\left[\left(D_{0}+m^{bare}\right)-\omega_{
      kin}\mathbf{D}^{2}-\omega_{spin}\boldsymbol{\sigma}\cdot\mathbf{B}\right]\psi_{h}\,,
\end{equation}
where the parametres $\omega_{kin}$ and $\omega_{spin}$ were computed
by applying the step-scaling method on the gauge field configurations
that include $n_f=2$ Wilson improved dynamical light quarks, and by
using SF technique. To that order in heavy quark expansion one can
write,
\begin{eqnarray}
  M_{B} & = &
  m^{bare}_b+E_{stat}+\omega_{kin}E_{kin}+\omega_{spin}E_{spin}\,,
\end{eqnarray} 
where $E_{stat}$ is determined from the correlation function of the
static heavy-light pseudoscalar/vector meson, while $E_{kin}$,
$E_{spin}$ are determined from time behaviour of the correlation
functions which include operator insertions of $\mathbf{D}^{2}$ and
$\boldsymbol{\sigma}\cdot\mathbf{B}$, respectively.

To determine $m_{b}^{bare}$, one interpolates
$M_{B}\left(m_b^{bare}\right)$ to reproduce $M_{B}^{exp}$, while
simultaneously extrapolating to the physical pion mass and to the
continuum limit.
They determined $m_b^{RGI}$ by studying the running of the quark mass
in the SF renormalization scheme and finally converted it to
$m_b^\MS(m_b)$ using the continuum perturbation theory.
Their result for the $b$ quark mass,
\begin{equation}
  m_b^\MS(m_b,\,n_f=5)=4.21(11)\,\GEV\,,
\end{equation}
was presented in Ref.~\cite{Bernardoni:2013xba}, which improved their
previous result based on quenched simulations~\cite{Della
  Morte:2006cb}. More importantly, $1/m_h$ corrections included in the
new study turned out to be very small, showing a reassuring
convergence of HQET at the scale of the $b$ quark mass,
and thus warranting the robustness of the approach.

\begin{figure}
  \includegraphics[width=1\textwidth]{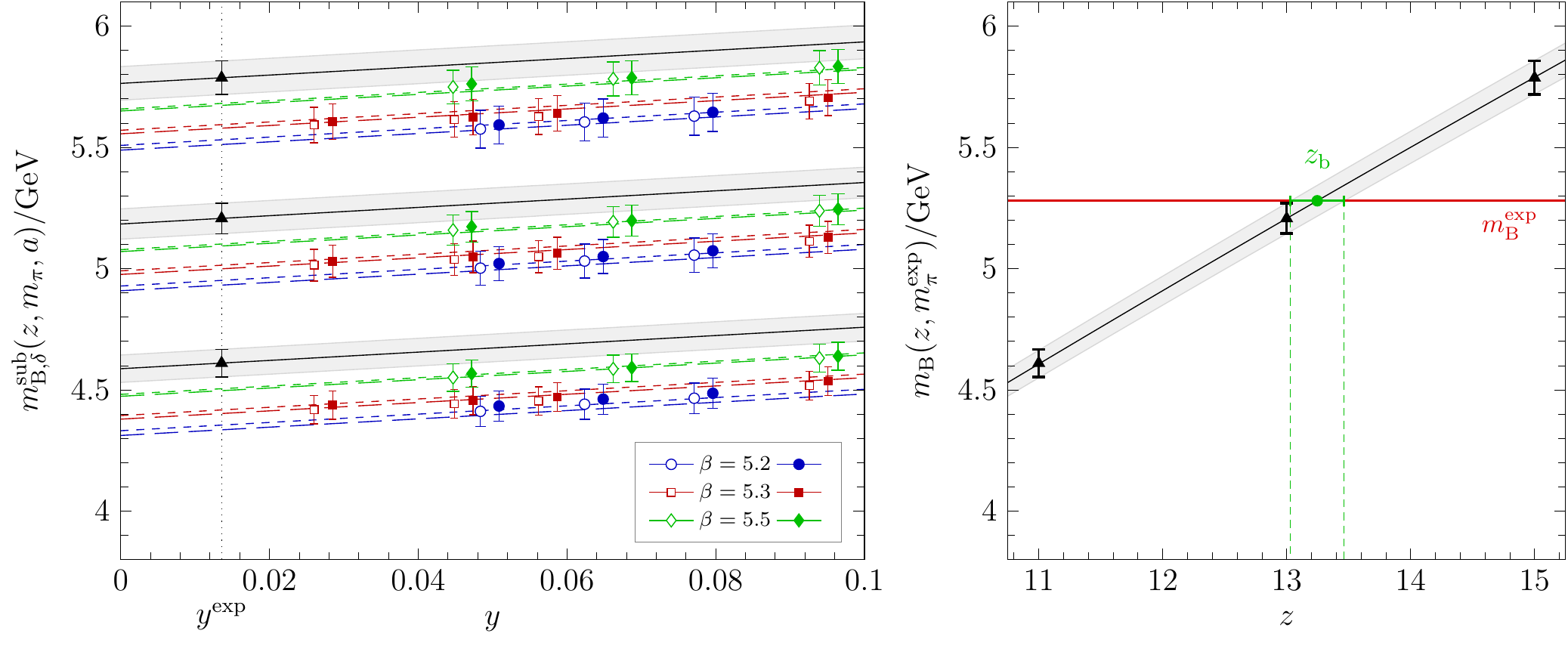}
  \caption{Chiral and continuum limit extrapolation (left panel) and
    interpolation to $b$ quark mass $z\propto m_b^{RGI}$ (right panel)
    of $M_B$ in HQET. Plots taken from
    Ref.~\cite{Bernardoni:2013xba}.}\label{alfig}
\end{figure}

\subsection{Ratio method}\label{ratiomet}

Finally, we discuss the method adopted by ETMC, in which the
observables used to fix the $b$ quark mass are computed at the scale
of the $b$ quark by interpolating between the results obtained around
the heavy charm quark mass and the static quark mass limit. To be able
to do so they adopted the so-called {\it ratio
  method}~\cite{Blossier:2009hg}, inspired by the Step-Scaling
approach proposed in Ref.~\cite{deDivitiis:2003iy}.
This method benefits from the heavy quark symmetry relation,
\begin{equation}
  \lim_{m_h\to\infty}\frac{M_{hl}}{m_{h}}=1\,\label{divergency},
\end{equation}
by considering a geometric series of masses
$m^{\left(0\right)}=m_{c}$, $m^{\left(1\right)}=\lambda m_{c}$,
\dots\,, $m^{\left(n\right)}=\lambda^{n}m_{c}\,$, from which one
computes ratios of $M_{hl}$ between two consecutive values of the
heavy quark mass,
\begin{equation}
  y\left(m_{h}^{\left(n\right)},\,\lambda;\,m_{l},\,a\right) =
  \lambda\frac{M_{hl}\left(m_{h}^{\left(n\right)};\,m_{l},\,a\right)}
              {M_{hl}\left(m_{h}^{\left(n\right)}/\lambda;\,m_{l},\,a\right)}\,.\label{ratio}
\end{equation}
Important advantages in computing these ratios on the lattice are a
large cancellation of cut-off effects, and a strong reduction of
statistical noise. Each ratio is extrapolated to the continuum limit
and to the physical Pion mass. The value of $M_{hl}(\lambda^n m_c)$
can be reconstructed by fitting the first $y^1\dots y^k$ ratios as a
function of $m_h=\lambda^k m_c$ up to values $k<n$ where the cut-off
effects are under control, and one can then easily compute
$M_{hl}(\lambda^n m_c)=M_{cl}y(\lambda m_c) y(\lambda^2 m_c) \dots
y(\lambda^n m_c)$ using the $y$ values from the fit ansatz. In this
fit a heavy quark symmetry relation, Eq.~(\ref{divergency}),
constrains the fit function and transforms the extrapolation to
interpolation.

With a given parameter $\lambda$, and in order to reproduce the
physical value $M_{B}$, one needs $j$ ratios, so that the quark mass
$m_{b}$ can be finally determined by reconstructing the corresponding
value $m_b=\lambda^j m_c$. In Ref.~\cite{Bussone:2014cha} the ETM
collaboration provided a preliminary result for $b$ quark mass,
\begin{equation}
  m_b^\MS(m_b) = 4.26(16)\,\GEV,
\end{equation}
from simulations with $n_f=2+1+1$ twisted mass quarks, thus improving
their previous result~\cite{Carrasco:2013zta}.

It must be stressed that the method relies substantially on the
cancellation of cut-off effects when computing ratio of an observable
at two close masses. This cancellation can and must be checked on a
case by case approach. In particular it has been shown to happen in
Twisted Mass, for a set of observables. In other regularisations or
for other observables the cut-off effects might change in a less
controlled way with the heavy quark mass, in which case the reduction
of cut-off effects would be only mild and the ratio method itself of
limited utility.

\section{$d$-$u$ quark mass difference}\label{ud_diff}

Isospin is a good symmetry of the strong interaction. Smallness of the
isospin breaking effects warrants reliability of the calculations
performed in the isospin symmetric limit of lattice QCD, in which it
is assumed that $m_u=m_d=m_l$, and charge are electrically neutral
($e=0$). On the other hand that same smallness makes difficult to
determine the mass difference $\delta m_{ud}=(m_d-m_u)/2$ itself. Far
from being a matter of purely conceptual relevance, determining
separately the mass of the light quarks answers deep theoretical
questions: for instance if the $u$ quark was massless, the
CP-violating term $\theta_{QCD}F\tilde{F}$ would not impact any
physical observable, which would then explain why the measured value
of $\theta_{QCD}$ is apparently so small (strong CP puzzle).

An updated review of the inclusion of QED into lattice QCD simulations
has been presented at this conference by A.~Portelli. In the following
we will cover only the aspects strictly related to the quark masses.
Including QED in the QCD Lagrangian complicates the pattern of
renormalisation of quark masses. Quarks of different charges receive
QED corrections that evolve in different ways under the
renormalisation group. For this reason, ratios of quark masses of
different charges are no longer RGI quantities (though the effect
induced by different running is largely negligible at the level of
current precision). Similarly the separation of the contribution of
QCD and QED to physical observables is a matter of convention, because
their sources (quark mass difference, QED corrections) get mixed up
under renormalisation. The exact separation of QCD and QED effects
requires an additional renormalisation condition, though a broad range
of sensible schemes can be considered to be equivalent at the level of
current precision.

Thanks to isospin and charge symmetries, the mass of neutral pion
receives corrections due to the breaking of isospin only at (highly
suppressed) $\mathcal{O}\left(e^2\delta m_{ud}\right)$. This makes its
experimental value $M_\pi^0=135\,\MEV$ appropriate to determine the
average mass of the light quarks $m_l$. The difference between the
neutral and charged Pion masses at leading order in the isospin
breaking is an $\mathcal{O}(e^2)$ effect, whereas the contribution due
to $\delta m_{ud}$ starts from $\mathcal{O}(\delta m_{ud}^2)$, and
thus cannot be efficiently used to determine $m_u$ and $m_d$
separately. The easiest solution is to consider to the difference of
masses of $K$ mesons. Such a strategy has been followed by three
different groups.

The RM123 collaboration included in a non-perturbative QCD framework
the leading order isospin breaking terms on the $n_f=2$ gauge
configurations generated by ETMC, and
obtained~\cite{deDivitiis:2013xla}
\begin{eqnarray}
  m_{u,\,d}^\MS(2\,{\rm GeV}) & = & \left\{2.40(15)(17),\,4.80(15)(17)\right\}\,\MEV\,,\\
  m_{u}/m_{d} & =& 0.50(2)(3)\,.
\end{eqnarray}

The BMW collaboration combined the determination of the ChPT Low
Energy Constant $B_{2}$~\cite{Durr:2013goa} with the Kaon meson mass
difference of~\cite{Borsanyi:2013lga}, to obtain the preliminary results
\begin{eqnarray}
  m_{u,\,d}^\MS(2\,{\rm GeV}) & = & \left\{2.29(6)(5),\,4.65(6)(5)\right\}\,\MEV\,,\\
  m_{u}/m_{d} & = & 0.49(1)(1)\,.
\end{eqnarray}

Finally, the MILC collaboration presented their preliminary results
for $m_u/m_d$ based on the update of the Kaon mass splitting
determined in Ref.~\cite{Basak:2013iw}. After combining that value
with the quark mass dependence found in the analysis of decay
constants (cf. contribution by J.~Komijani), they
obtain~\cite{Basak:2014vca}
\begin{equation}
  m_u/m_d = 0.4482 (48)_{\rm stat} ({}^{+\phantom{0}21}_{-115})_{a^2} (1)_{\rm FV_{QCD}} (165)_{\rm EM}\,,
\end{equation}
where the dominant source of error comes from electromagnetism.

The precision reached by the three methods, based on largely different
approaches, and the general agreement of the results, can be viewed as
progress of lattice QCD. Moreover, all three results completely rule
out the possibility that the $u$ quark is massless.

\section{Conclusions}\label{conclusions}

The determination of the quark masses is of great theoretical and
phenomenological importance:
\begin{itemize}
  \setlength\itemsep{0.2em}
\item once determined, they become input parameters for other calculations in
  QCD;
\item they help restraining the (representations of) gauge groups of
  theories aiming at the Grand Unification of fundamental
  interactions;
\item more generally, they provide input and constraints to any theory
  of flavour;
\item compatibility of the determinations obtained by different lattice
  groups, is also a very significant check of universality of the
  continuum limit of lattice QCD.
\end{itemize}
Lattice QCD offers a well defined framework in which to determine
quark masses non-perturbatively. Several approaches can be used to
define renormalised quark masses and to match them to commonly used
schemes such as $\MS$. Ratios of the masses of different quarks, being
RGI quantities, can be and are computed very accurately on the
lattice, and can be very useful when comparing the lattice and
non-lattice approaches in treating non-perturbative QCD.

Thanks to the recent theoretical and numerical developments, we are
nowadays able to compute quite accurately the $b$ quark mass and the
difference between $u$ and $d$ quark masses. In that way we are
completing the picture of determination of quark masses by means
of lattice QCD.

The results concerning the light quark masses discussed here are in
good agreement with those extensively discussed in the last FLAG
report~\cite{Aoki:2013ldr}. Awaiting an update of the FLAG review
that will also include a detailed discussion of heavy quarks, it is
useful to quote the average value of $m_c$ obtained from the two
$n_f=2+1+1$ simulations discussed in Sec.~\ref{reno_massive_qua} and
Sec.~\ref{moments},
\begin{equation}
  m_c^\MS(m_c,\,n_f=4)=1.27(1)\,\GEV\,.
\end{equation}
However it would be desirable to have at least one more lattice QCD
estimate of $m_c$ with $n_f=2+1+1$ in order to clarify a discrepancy
between the results obtained by two collaborations.

As for the average $b$ quark mass of the $n_f=2+1+1$ determinations
discussed in Sec.~\ref{moments} and Sec.~\ref{ratiomet} we obtain,
\begin{equation}
  m_b^\MS(m_b,\,n_f=5)=4.17(5)\,\GEV\,.
\end{equation}

\section{Acknowledgements}
I warmly thanks the organizers of the conference for the invitation to
present this review talk, and all the speakers that sent me
contributions. I particularly thanks D.~Becrievic, V.~Lubicz,
C.~Sachrajda for the precious comments on the manuscript.  This work
has been supported by the European Reasearch Council under the
European Union's Seventh Framework Progamme (FP7/2007-2013) / ERC
Grant agreement 27975.

\end{document}